\documentclass{article}
\usepackage{spconf,amsmath,graphicx}

\usepackage{amsmath,amssymb,color,amsmath, graphicx, float, caption, subcaption, mathrsfs,color} \usepackage[ruled]{algorithm2e}
\usepackage{bm}
\usepackage{epstopdf }
\usepackage{cite}

\usepackage{scalerel,stackengine}
\stackMath
\newcommand\reallywidehat[1]{%
\savestack{\tmpbox}{\stretchto{%
  \scaleto{%
    \scalerel*[\widthof{\ensuremath{#1}}]{\kern-.6pt\bigwedge\kern-.6pt}%
    {\rule[-\textheight/2]{1ex}{\textheight}}
  }{\textheight}%
}{0.5ex}}%
\stackon[1pt]{#1}{\tmpbox}%
}
\usepackage{selinput}
\SelectInputMappings{%
  eacute={é},
}
\renewcommand{\thetable}{S\arabic{table}}
\usepackage{multirow}

\newcounter{para}

\def\A{\bm{A}}

\def\I{\bm{I}}

\def\N{\bm{N}}

\def\R{\bm{R}}

\def\W{\bm{W}}

\def\e{\bm{e}}

\def\g{\bm{g}}

\def\x{\bm{x}}
\def\y{\bm{y}}
\def\z{\bm{z}}

\def\balpha{\boldsymbol{\alpha}}

\def\argmin#1{\underset{#1}{\textrm{argmin}}}

\def\CRLB_pe{\text{CRLB}}
\def\CRLB_pe{\text{CRLB}(\mathbf{p}_{e})}

\def\N02{\frac{N_0}{2}}

\def\0{\mathbf{0}}
\def\1{\mathbf{1}}

\SetKwInput{kwEvaluate}{Evaluate}
\SetKwInput{kwSort}{Sort}
\SetKwInput{kwInput}{Input}
\SetKwInput{kwOutput}{Output}
\SetKwInput{kwInitialize}{Initialize}
\SetKwInput{kwParameters}{Params.}
\SetKwInput{kwDefaults}{Default init.}

\title{An Interpretation of Regularization by Denoising and its Application with The Back-Projected Fidelity Term}
\name{Einav Yogev-Ofer, Tom Tirer, and Raja Giryes}
\address{School of Electrical Engineering\\
Tel Aviv University\\}
\begin{document}
%
\maketitle
%


\begin{abstract}

The vast majority of image recovery tasks are ill-posed problems. As such, methods that are based on optimization use cost functions that consist of both fidelity and prior (regularization) terms. 
A recent line of works imposes the prior by the Regularization by Denoising (RED) approach, which exploits the good performance of existing image denoising engines. Yet, the relation of RED to explicit prior terms is still not well understood, as previous work requires too strong assumptions on the denoisers.
In this paper, we make two contributions.
First, we show that the RED gradient can be seen as a (sub)gradient of a prior function---but taken at a denoised version of the point. 
As RED is typically applied with a relatively small noise level, this interpretation indicates a similarity between RED and traditional gradients. 
This leads to our second contribution: We propose to combine RED with the Back-Projection (BP) fidelity term rather than the common Least Squares (LS) term that is used in previous works. 
We show that the advantages of BP over LS for image deblurring and super-resolution, which have been demonstrated for traditional gradients, carry on to the RED approach.

\end{abstract}

%
%
\section{Introduction}
\label{sec:intro}

Image recovery tasks aim to restore an original image from its observed degraded version. 
The observed image can be degraded in many ways, such as blurring, subsampling (lower resolution), noise addition, or all together. In many tasks, the observed image can be described by a linear expression: 
\begin{align}
\label{eq:linear_eq}
    \y = \A\x_* + \e,
\end{align}
Where $\x_*\in \mathbb{R}^n$ is the original image, $\y \in \mathbb{R}^m$ is the degraded observed image, $\e \in \mathbb{R}^m$ is a noise vector and $\A \in \mathbb{R}^{m \times n}$ is the degradation matrix. For example, in the deblurring task $\A$ 
 represents the blurring operation, 
 and in the super-resolution task $\A$ is an
 operator that represents blurring (typically, anti-aliasing filtering) followed by subsampling.

Most of the methods for recovering $\x_*$ from the observed $\y$
are based on optimization. They involve minimization of a cost function that is composed of a data fidelity term and a prior term
\begin{align}
\label{eq:cost_function}
    f(\x) = \ell(\x) + \lambda s(\x),
\end{align}
where $\ell(\cdot)$ is the fidelity term, $s(\cdot)$ is the prior term and $\lambda$ is a positive parameter that controls the level of regularization. The fidelity term encourages 
complying with the observation model \eqref{eq:linear_eq}, 
while the prior represents assumptions on the original image, 
and is inevitable due to the ill-posedness nature of image reconstruction tasks.

Over the years, different prior functions have been proposed, such as $\ell_1$-norm of the wavelet coefficients vector \cite{donoho1994ideal,donoho1995noising} and the Total Variation (TV) prior \cite{Rudin92TV}.
Interestingly, there are also very successful priors, such as BM3D \cite{Dabov07BM3D}, that are based on a series of operations rather than on an explicit prior function.
Traditionally, a different algorithm has been designed for each task (structure of $\A$ in \eqref{eq:linear_eq}) and each prior $s(\cdot)$.
However, it has been suggested in \cite{venkatakrishnan2013plug} to exploit the good  performance of existing image denoising for solving tasks other then denoising. This ``plug-and-play denoisers" concept encourages one to use a denoiser for imposing the prior even if it not explicitly clear what is the function $s(\cdot)$ that is associated with this denoiser (as in the case of BM3D).
Follow-up papers of \cite{venkatakrishnan2013plug} include \cite{sreehari2016plug,metzler2016denoising,RED,bigdeli2017deep,zhang2017learning,tirer2018image,buzzard2018plug}, and many more.

A popular line of works (see, e.g., \cite{RED,bigdeli2017deep,reehorst2018regularization,mataev2019deepred,sun2020block,sun2020async,cohen2020regularization}) that follows this plug-and-play
concept is based on the
Regularization by Denoising (RED) approach. 
The original RED paper \cite{RED} proposed a gradient-based regularization of inverse problems using an off-the-shelf Gaussian denoiser $\mathcal{D}(\cdot;\sigma)$ ($\sigma$ is the noise level of the denoiser, which is not necessarily the noise level in $\y$).
Specifically, within optimization schemes, 
it is proposed to replace $\nabla s(\x)$ with the ``RED gradient"
\begin{align}
\label{eq:red_grad}
    \g_{\text{\tiny{RED}}}(\x;\mathcal{D}) := \x - \mathcal{D}(\x;\sigma).
\end{align}
The paper \cite{RED} showed that under strong assumptions on the denoiser, 
which as pointed out in \cite{reehorst2018regularization} include also a symmetric Jacobian condition, 
the expression in \eqref{eq:red_grad} is the gradient of the prior term 
\begin{align}
\label{eq:red_prior}
    s_{\text{\cite{RED}}}(\x) = \frac{1}{2}\x^T(\x - \mathcal{D}(\x;\sigma)).
\end{align}
However, as shown in \cite{reehorst2018regularization}, these assumptions do not hold for widely used denoisers. 
Therefore, using RED may be better understood as a {\em prior that is imposed directly on the gradient} of a given optimization scheme rather than on the optimization objective \cite{reehorst2018regularization,bigdeli2017deep}. 


In this paper, we show that the RED gradient can be seen as a (sub)gradient of a prior function (beyond the class of functions that can be expressed as \eqref{eq:red_prior}), but taken at a denoised version of the point. 
As RED is typically applied with a relatively small noise level, this interpretation indicates a similarity between the RED gradient and traditional gradients. 
Following this interpretation, 
we propose to combine RED with the Back-Projection (BP) fidelity term \cite{Tirer2019BackProjectionBF} rather than with the common Least Squares (LS) fidelity term that is used in previous works \cite{RED,bigdeli2017deep,reehorst2018regularization,mataev2019deepred,sun2020block,sun2020async,cohen2020regularization}. 
We show that the advantages of BP over LS for image deblurring and super-resolution, which have been demonstrated for traditional gradients \cite{tirer2018image,tirer2019super,Tirer2019BackProjectionBF,tirer2020convergence,zukerman2020bp}, carry on to the RED approach.


\section{An Interpretation for RED}
\label{sec:Interpretation}



We start with providing an interpretation for RED that may provide an additional insight on its applicability and success.  

Note that a Gaussian denoiser $\mathcal{D}(\x;\sigma)$ of an image $\x$, under the prior $s(\cdot)$ and noise level $\sigma>0$, can be expressed as the minimizer of the following optimization problem
\begin{eqnarray}
\label{eq:cost_denoising}
    \mathcal{D}(\x;\sigma) = 
    \argmin{\z} \,  \frac{1}{2\sigma^2}\| \z-\x \|_2^2 + s(\z).
\end{eqnarray}
For a proper lower-semicontinuous convex function $s(\cdot)$ this minimizer is unique, and in the optimization literature 
it is often referred to as the proximal mapping of $\sigma^2 s(\cdot)$ at $\x$.
For a convex $s(\cdot)$, the first-order optimality condition of $\hat{\x} =\mathcal{D}(\x;\sigma)$ implies 
that 
\begin{align}
\label{eq:cost_denoising_opt}
    \bm{0} \in \hat{\x} - \x + \sigma^2 \partial s(\hat{\x}),
\end{align}
where $\partial s(\hat{\x}):=\{\g : s(\z) \geq s(\hat{\x})+\g^T(\z-\hat{\x}), \forall \z \}$ is the subdifferential (the set of the subgradients) of $s(\cdot)$ at $\hat{\x}$. 
Thus, $\x-\hat{\x}=\x-\mathcal{D}(\x;\sigma) \in \sigma^2 \partial s(\hat{\x})$ and the gradient of RED (see \eqref{eq:red_grad}) obeys
\begin{align}
\label{eq:red_grad_ourInterp}
    \g_{\text{\tiny{RED}}}(\x;\mathcal{D}) \in \sigma^2 \partial s(\mathcal{D}(\x;\sigma)).
\end{align}
If $s(\cdot)$ is a smooth (i.e. continuously differentiable), then $\partial s(\hat{\x})=\nabla s(\hat{\x})$ (a singleton), so
\begin{align}
\label{eq:red_grad_ourInterp_smooth}
    \g_{\text{\tiny{RED}}}(\x;\mathcal{D}) = \sigma^2 \nabla s(\mathcal{D}(\x;\sigma)).
\end{align}

As the factor $\sigma^2$ can be absorbed in the hyper-parameter $\lambda$ (cf. \eqref{eq:cost_function}), we see that for a general prior $s(\cdot)$,
the RED gradient $\g_{\text{\tiny{RED}}}(\x)$\footnote{For brevity, from now on we omit the dependency on $\mathcal{D}$.} can be seen as a gradient of $s(\cdot)$ at a cleaner point, i.e., after denoising using the prior $s(\cdot)$. 
Therefore, $\g_{\text{\tiny{RED}}}(\x)$ enhances the effect of the prior compared to using the gradient $\nabla s(\x)$ at the current point. Let us illustrate this relationship in two concrete examples.


{\bf Tikhonov regularization.}
For the simple Tikhonov regularization prior $s(\x)=\frac{1}{2}\|\R\x\|_2^2$, 
we have that 
$\nabla s(\x)=\R^T\R\x$ and $\mathcal{D}(\x;\sigma)=(\sigma^2 \R^T\R+\I_n)^{-1}\x$.
Thus, from \eqref{eq:red_grad_ourInterp_smooth} we get
\begin{eqnarray}
\label{eq:red_grad_tikhonov_ourInterp}
    \g_{\text{\tiny{RED}}}(\x) &=& \sigma^2 \nabla s(\mathcal{D}(\x;\sigma)) \\ \nonumber
    &=& \sigma^2 \R^T\R (\sigma^2 \R^T\R+\I_n)^{-1}\x.
\end{eqnarray}
As a sanity check, notice that this result indeed coincides with the definition in \eqref{eq:red_grad}, i.e.,
\begin{align}
\label{eq:red_grad_tikhonov}
 \g_{\text{\tiny{RED}}}(\x) &= \sigma^2 \R^T\R (\I_n + \sigma^2 \R^T\R)^{-1} \x   \nonumber \\
    &= \x - ( \I_n - \sigma^2 \R^T\R (\I_n + \sigma^2 \R^T\R)^{-1}  ) \x \nonumber \\
    &=     \x - (\sigma^2 \R^T\R+\I_n)^{-1}\x \nonumber \\
    &= \x - \mathcal{D}(\x;\sigma),
\end{align}
where the second equality uses the Woodbury identity.

Since the outer $\sigma^2$ in \eqref{eq:red_grad_tikhonov_ourInterp} can be absorbed in the hyper-parameter $\lambda$, 
we are left with comparing $\R^T\R$ (that appears in $\nabla s(\x)$) and $\R^T\R (\sigma^2 \R^T\R+\I_n)^{-1}$ (that appears in $\g_{\text{\tiny{RED}}}(\x)$ in  \eqref{eq:red_grad_tikhonov_ourInterp}).
Clearly $\R^T\R (\sigma^2 \R^T\R+\I_n)^{-1} \leq \R^T\R$ (in the sense that $\R^T\R-\R^T\R (\sigma^2 \R^T\R+\I_n)^{-1}$ is positive semi-definite). 
This demonstrates the enhanced effect of the prior that is inherent in $\g_{\text{\tiny{RED}}}(\x)$. Yet, observe that for a small values of $\sigma$ (as typically used in RED) the directions of $\nabla s(\x)$ and $\g_{\text{\tiny{RED}}}(\x)$ will be close.

\begin{figure*}
    \centering
    \hfill    
    \begin{subfigure}[b]{0.245\textwidth}
        \centering
        \includegraphics[width=\textwidth]{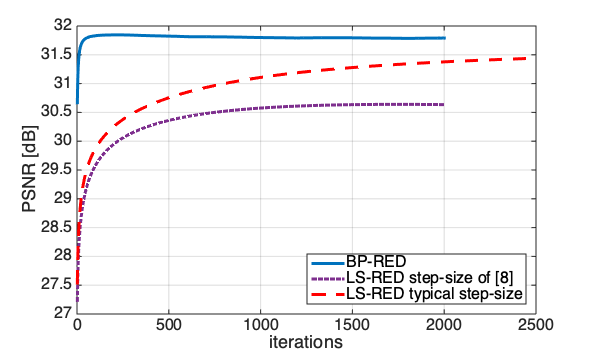}
        \captionsetup{width=.9\linewidth}
        \caption{Gaussian blur kernel, $\sigma_e=\sqrt{0.3}$, using BM3D prior}    
        \label{fig:deblurring_gaussian_sc1_bm3d_iterations}
    \end{subfigure}
    \hfill   
    \begin{subfigure}[b]{0.245\textwidth}
        \centering
        \includegraphics[width=\textwidth]{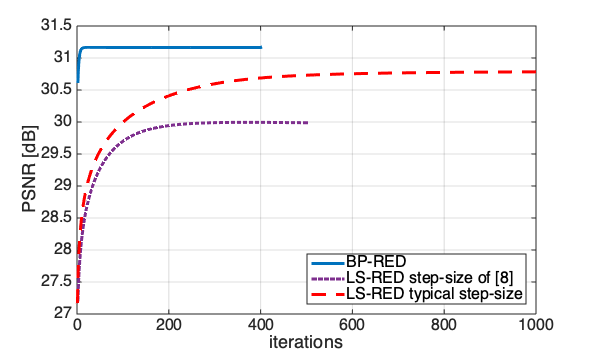}
        \captionsetup{width=.9\linewidth}
        \caption{Gaussian blur kernel, $\sigma_e=\sqrt{0.3}$, using TV prior}    
        \label{fig:deblurring_gaussian_sc1_tv_iterations}
    \end{subfigure}
    \hfill  
    \begin{subfigure}[b]{0.245\textwidth}
        \centering
        \includegraphics[width=\textwidth]{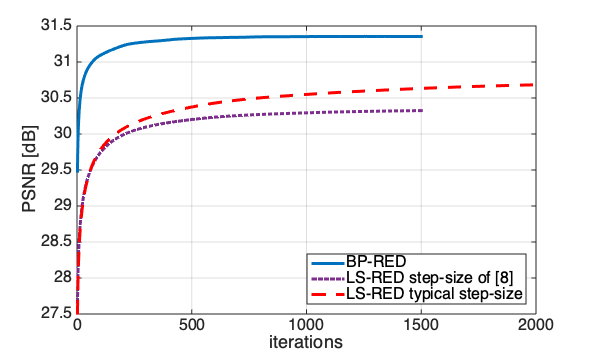}
        \captionsetup{width=.9\linewidth}
        \caption{Gaussian blur kernel, $\sigma_e=\sqrt{2}$, using BM3D prior}    
        \label{fig:deblurring_gaussian_sc2_bm3d_iterations}
    \end{subfigure}
    \hfill  
    \begin{subfigure}[b]{0.245\textwidth}
        \centering
        \includegraphics[width=\textwidth]{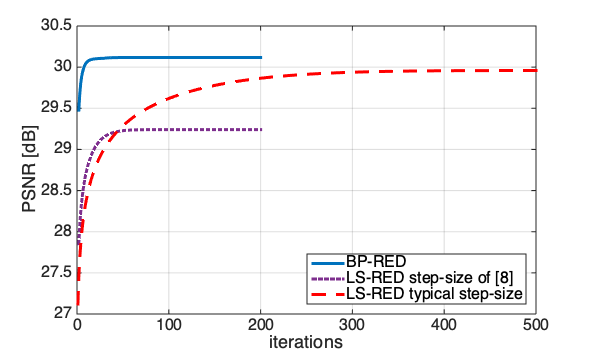}
        \captionsetup{width=.9\linewidth}
        \caption{Gaussian blur kernel, $\sigma_e=\sqrt{2}$, using TV prior}    
        \label{fig:deblurring_gaussian_sc2_tv_iterations}
    \end{subfigure}
    \hfill      
    \begin{subfigure}[b]{0.245\textwidth}
        \centering
        \includegraphics[width=\textwidth]{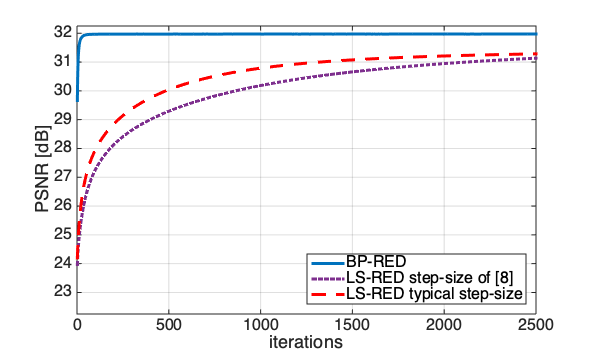}
        \captionsetup{width=.9\linewidth}
        \caption{Uniform blur kernel, $\sigma_e=\sqrt{0.3}$, using BM3D}
        \label{fig:deblurring_uniform_sc1_bm3d_iteratins}
    \end{subfigure}
    \hfill  
    \begin{subfigure}[b]{0.245\textwidth}
        \centering
        \includegraphics[width=\textwidth]{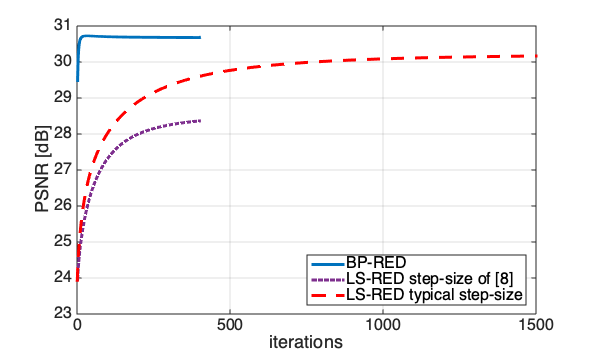}
        \captionsetup{width=.9\linewidth}
        \caption{Uniform blur kernel, $\sigma_e=\sqrt{0.3}$, using TV prior}   
        \label{fig:deblurring_uniform_sc1_tv_iterations}
    \end{subfigure}
    \hfill  
    \begin{subfigure}[b]{0.245\textwidth}
        \centering
        \includegraphics[width=\textwidth]{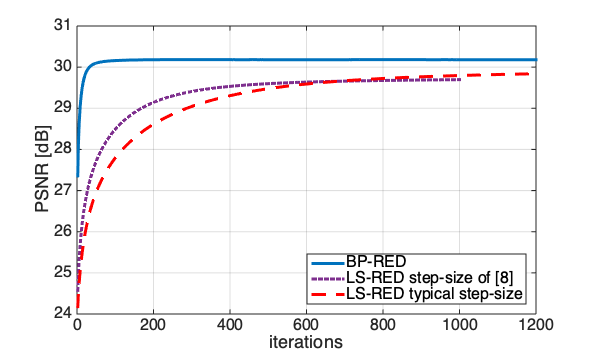}
        \captionsetup{width=.9\linewidth}
        \caption{Uniform blur kernel, $\sigma_e=\sqrt{2}$, using BM3D prior}    
        \label{fig:deblurring_uniform_sc2_bm3d_iterations}
    \end{subfigure}
    \hfill    
    \begin{subfigure}[b]{0.245\textwidth}
        \centering
        \includegraphics[width=\textwidth]{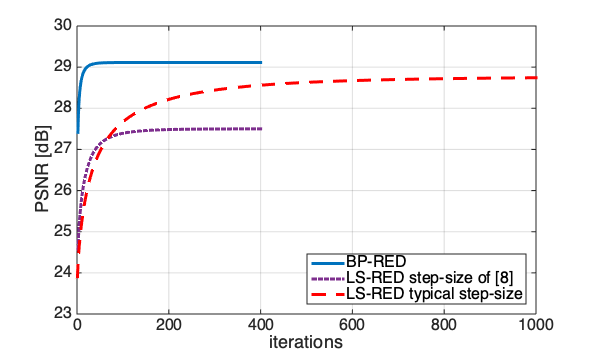}
        \captionsetup{width=.9\linewidth}
        \caption{Uniform blur kernel, $\sigma_e=\sqrt{2}$, using TV prior}    
        \label{fig:deblurring_uniform_tv_sc2_iteratins}
    \end{subfigure}
    \hfill  
        \caption{Deblurring results: PSNR (averaged over 8 test images) vs.~iteration number, for LS-RED (red) and BP-RED (blue).}
    \label{fig:delurring_iterations}
\end{figure*}

{\bf $\bm{\ell_1}$-norm prior.}
Let us observe the relation between $\partial s(\x)$ and $\g_{\text{\tiny{RED}}}(\x)$ 
for the (nonsmooth) prior $s(\x)=\|\W^T\x\|_1$, where $\W$ is an orthonormal basis (i.e., $\W^T\W=\W\W^T=\I_n$).\footnote{Note that this prior cannot be expressed in the form of \eqref{eq:red_prior} (see \cite{RED,reehorst2018regularization}).}  
For example, $\W$ can be an orthonormal wavelet basis under which the coefficients vector $\balpha=\W^T\x$ is sparse. 
In this case, the subdifferential $\partial s(\x)$ is given by 
\begin{align}
\label{eq:subdif_Well1}
    \partial s(\x) &= \W \partial \|\balpha\|_1 \big|_{\balpha=\W^T\x} \nonumber \\
    &= \W \left \{ \z : \begin{cases}
        z_i=1, & [\W^T\x]_i > 0\\
        z_i \in [-1,1], & [\W^T\x]_i = 0\\
        z_i=-1, & [\W^T\x]_i < 0
        \end{cases} \right \}.    
\end{align}

For the $\ell_1$-norm prior, 
note that 
\eqref{eq:cost_denoising} turns into the soft-thresholding denoiser \cite{donoho1994ideal,donoho1995noising}. Namely,
\begin{align}
\label{eq:cost_denoising_ell1}
    \mathcal{D}(\x;\sigma) &= \argmin{\z} \,  \frac{1}{2}\| \z-\x \|_2^2 + \sigma^2 \|\W^T\z\|_1 \nonumber \\
    &= \W \left (\argmin{\tilde{\balpha}} \,  \frac{1}{2}\| \tilde{\balpha} - \W^T \x \|_2^2 + \sigma^2 \| \tilde{\balpha} \|_1 \right ) \nonumber \\
    &= \W \mathcal{T}_{\sigma^2}(\W^T \x),
\end{align}
where $[\mathcal{T}_{\sigma^2}(\balpha)]_i=\mathrm{sign}(\alpha_i)\mathrm{max}(|\alpha_i|-\sigma^2,0)$.
The subdifferential $\sigma^2 \partial s(\mathcal{D}(\x;\sigma))$ is then given by
\begin{align}
\label{eq:red_grad_ell1_verify}
    &\sigma^2 \partial s(\mathcal{D}(\x;\sigma)) = \sigma^2 \W \partial \|\balpha\|_1 \big|_{\balpha=\W^T\mathcal{D}(\x;\sigma)} \nonumber \\
    &\hspace{5mm} =  \sigma^2 \W \partial \|\balpha\|_1 \big|_{\balpha= \mathcal{T}_{\sigma^2}(\W^T \x) } \nonumber \\
    &\hspace{5mm} = \sigma^2 \W \left \{ \z : \begin{cases}
    z_i=1, & [\mathcal{T}_{\sigma^2}(\W^T\x)]_i > 0\\
    z_i \in [-1,1], & [\mathcal{T}_{\sigma^2}(\W^T\x)]_i = 0\\
    z_i=-1, & [\mathcal{T}_{\sigma^2}(\W^T\x)]_i < 0
    \end{cases} \right \}.
\end{align}
Comparing \eqref{eq:red_grad_ell1_verify} and \eqref{eq:subdif_Well1}, the enhanced effect of the nonsmoothness of the prior 
is clear:  
$\partial s(\mathcal{D}(\x;\sigma)) \supseteq \partial s(\x)$.

Let us verify \eqref{eq:red_grad_ourInterp} for the RED gradient defined in \eqref{eq:red_grad} 
\begin{align}
\label{eq:red_grad_ell1}
    \g_{\text{\tiny{RED}}}(\x) &= \x - \mathcal{D}(\x;\sigma) = \x - \W \mathcal{T}_{\sigma^2}(\W^T \x)  \nonumber \\
    &= \sigma^2 \W \frac{1}{\sigma^2} \left ( \W^T \x - \mathcal{T}_{\sigma^2}(\W^T \x) \right )=:\sigma^2 \W \z_{\text{\tiny{RED}}}.
\end{align}
If $[\mathcal{T}_{\sigma^2}(\W^T\x)]_i > 0$, we have 
\begin{align}
\label{eq:red_grad_ell1_verify222}
    [\z_{\text{\tiny{RED}}}]_i = \frac{1}{\sigma^2}([\W^T \x]_i - ([\W^T \x]_i - \sigma^2)) = 1. 
\end{align}
Similarly, if $[\mathcal{T}_{\sigma^2}(\W^T\x)]_i < 0$, we have 
$[\z_{\text{\tiny{RED}}}]_i = -1$.
Lastly, if $[\mathcal{T}_{\sigma^2}(\W^T\x)]_i = 0$, or equivalently $-\sigma^2 \leq [\W^T \x]_i \leq \sigma^2$, we have 
    $[\z_{\text{\tiny{RED}}}]_i = \frac{1}{\sigma^2} [\W^T \x]_i \in [-1,1]$.
Thus, $\g_{\text{\tiny{RED}}}(\x) \in \sigma^2 \partial s(\mathcal{D}(\x;\sigma))$.

We believe that choosing $\g_{\text{\tiny{RED}}}(\x)$ and not other subgradients that are in \eqref{eq:red_grad_ell1_verify} is good, in the sense that it is closer (in angle) to subgradients from \eqref{eq:subdif_Well1},
and performs gradient-based restoration similar to subgradients from \eqref{eq:subdif_Well1} and better than other subgradients that are in \eqref{eq:red_grad_ell1_verify}. We defer proving this to future research.

\begin{figure*}
    \centering
    \hfill      
    \begin{subfigure}[b]{0.245\textwidth}
        \centering
        \includegraphics[width=\textwidth]{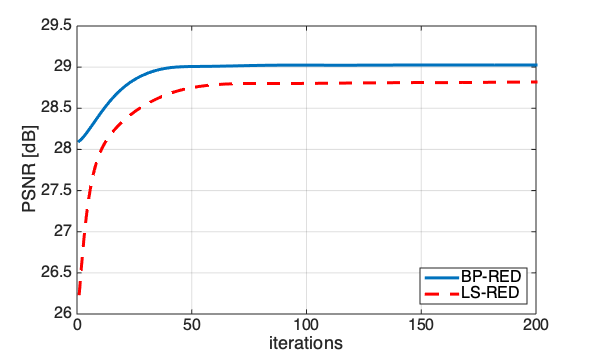}
        \captionsetup{width=.9\linewidth}
        \caption{SRx3 for Gauss.~kernel with std 1.6, using BM3D prior}
        \label{fig:SR_gaussian_scale3_BM3D_no_noise_iterations}
    \end{subfigure}
    \hfill  
    \begin{subfigure}[b]{0.245\textwidth}
        \centering
        \includegraphics[width=\textwidth]{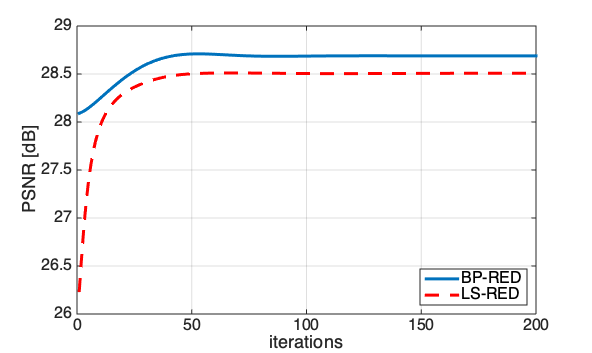}
        \captionsetup{width=.9\linewidth}
        \caption{SRx3 for Gauss.~kernel with std 1.6, using TV prior}    
        \label{fig:SR_gaussian_scale3_TV_no_noise_iterations}
    \end{subfigure}
    \hfill  
    \begin{subfigure}[b]{0.245\textwidth}
        \centering
        \includegraphics[width=\textwidth]{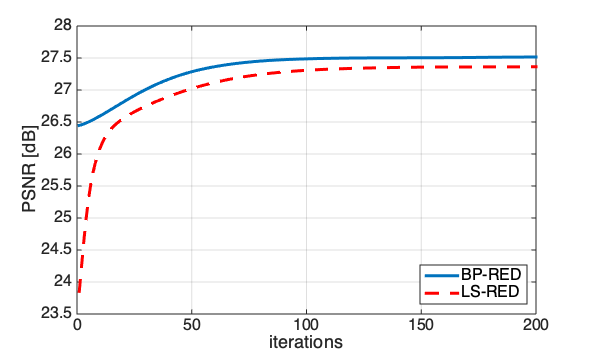}
        \captionsetup{width=.9\linewidth}
        \caption{SRx4 for Gauss.~kernel with std 2.2, using BM3D prior}    
        \label{fig:sr_gaussian_scale4_bm3d_iterations}
    \end{subfigure}
    \hfill    
    \begin{subfigure}[b]{0.245\textwidth}
        \centering
        \includegraphics[width=\textwidth]{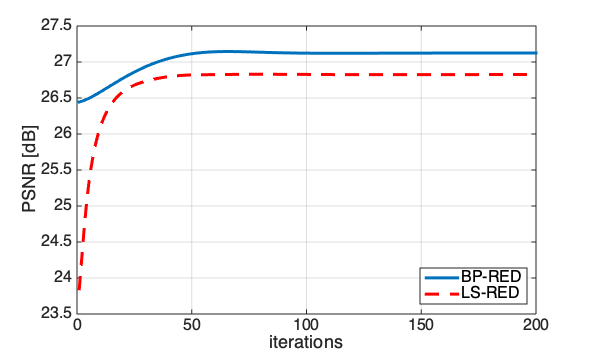}
        \captionsetup{width=.9\linewidth}
        \caption{SRx4 for Gauss.~kernel with std 2.2, using TV prior}    
        \label{fig:sr_gaussian_scale4_tv_iterations}
    \end{subfigure}
    \hfill  
    \caption{SR results: PSNR (averaged over 8 test images) vs.~iteration number, for LS-RED (red) and BP-RED (blue).}    
    \label{fig:sr_iterations}
\end{figure*}

\section{The Proposed Method: BP-RED}
\label{sec:method}

In the previous section we saw that the RED gradient differs from a traditional (sub)gradient of $s(\x)$ only in the fact that it is computed in a denoised version of $\x$ (associated with the prior $s(\x)$). 
As RED is typically applied with a relatively small noise level, this indicates a similarity between the RED gradient and traditional gradients,
which 
implies that approaches that are useful for traditional gradient descent are likely to be beneficial also for RED.
Following this perspective, we propose to incorporate the RED approach with the Back-Projection (BP) fidelity term \cite{Tirer2019BackProjectionBF}.

Previous works on RED use the LS term $\ell_{LS}(\x)=\frac{1}{2}\|\y-\A\x\|_2^2$ as their fidelity term. 
Formally, many of them are based on the following ``gradient descent" reconstruction scheme
\begin{eqnarray}
\label{eq:red_LS}
    \x_{k+1} &=& \x_{k} - \mu \left ( \nabla \ell_{LS}(\x_k) + \lambda \g_{\text{\tiny{RED}}}(\x_k)    \right )  \\ \nonumber
    &=& \x_{k} - \mu \left ( \A^T(\A\x_k-\y) + \lambda ( \x_k - \mathcal{D}(\x_k;\sigma) ) \right ),
\end{eqnarray}
where $\mu$ is the step-size.
Recent works on ill-posed linear inverse problems \cite{tirer2018image,tirer2019super,Tirer2019BackProjectionBF,tirer2020convergence,zukerman2020bp} have shown the benefits of using the BP fidelity term $\ell_{BP}(\x)=\frac{1}{2}\|\A^{\dagger}(\y-\A\x)\|_2^2$, where $\A^{\dagger}:=\A^T(\A\A^T)^{-1}$ is the pseudoinverse of 
$\A$, instead of LS.
Following this approach, we propose to combine the RED prior with the BP term. Thus, we propose using
\begin{align}
\label{eq:red_BP}
    \x_{k+1} &= \x_{k} - \mu \left ( \nabla \ell_{BP}(\x_k) + \lambda \g_{\text{\tiny{RED}}}(\x_k)    \right )  \\ \nonumber
    &= \x_{k} - \mu \left ( \A^\dagger(\A\x_k-\y) + \lambda ( \x_k - \mathcal{D}(\x_k;\sigma) )  \right ).
\end{align}
Note that in important tasks, such as deblurring and super-resolution, the operator $\A^\dagger$ can be implemented efficiently using FFT or conjugate gradients (CG) with no need for matrix inversion, and the computational complexity of \eqref{eq:red_BP} is similar to \eqref{eq:red_LS}. For sophisticated denoisers (e.g., TV \cite{Rudin92TV}, BM3D \cite{Dabov07BM3D}, or deep CNNs) the complexity of both \eqref{eq:red_LS} and \eqref{eq:red_BP} are dominated by the complexity of the denoising operation $\mathcal{D}(\cdot;\sigma)$.

\section{Experiments}
\label{sec:exp}

In this section, we concentrate on two image recovery tasks: deblurring and super-resolution,
which are performed on eight classical test images: cameraman, house, peppers, Lena, Barbara, boat, hill and couple.
%
We examine the performance of the two gradient-based methods in \eqref{eq:red_LS} and \eqref{eq:red_BP}, denoted by LS-RED and BP-RED, respectively. 
%
%
As the off-the-shelf denoisers, we use the (convex) isotropic total-variation (TV)\cite{Rudin92TV} and the (non-convex) BM3D\cite{Dabov07BM3D}.  

\subsection{Deblurring}\par

In the deblurring task, the linear operator $\A$ represents filtering with some blur kernel. 
We examine two common 
blur kernels: $9 \times 9$ uniform kernel and $9 \times 9$ Gaussian kernel with std of 1.6 (both are normalized to unit sum),  
and two levels of Gaussian noise: $\sigma_e=\sqrt{0.3}$ and $\sigma_e=\sqrt{2}$.


Both LS-RED and BP-RED have similar per-iteration complexity, which is dominated by the complexity of the denoisers, since
the operators $\A, \A^T$ and $\A^\dagger$ are implemented very efficiently using the Fast Fourier Transform (FFT) (see \cite{tirer2018image}). 
In the FFT implementation of $\A^\dagger$, we regularize the denominator by $0.01\sigma_e^2$. 
We initialize both methods with $\x_0=\y$. 
We use the typical step-size of 1 over the Lipschitz constant of $\nabla \ell(\cdot)$ for both methods, i.e., $\mu=1/||\A^\dagger\A|| = 1$ for BP-RED, and $\mu=1/||\A^T\A||$, computed by the power method, for LS-RED. For LS-RED, we also present results for step-size of $\mu = 2/(\frac{1}{\sigma_e^2} + \lambda)$ as used in \cite{RED}.
The two hyper-parameters of LS-RED and BP-RED (namely, the regularization level $\lambda$ and the noise level $\sigma$ that is used in the denoiser) are tuned uniformly for each scenario by choosing the best $\lambda$ from a dense grid in $(0.005,2.5)$ and the best $\sigma$ from a dense grid in $(0.5,20)$. 


The restoration performance of the two methods for all the experiments is presented in Fig.~\ref{fig:delurring_iterations}. The graphs show the Peak Signal to Noise Ratio (PSNR) metric (averaged over all the test images) vs.~the iteration number. One can see that using the BP-RED method achieves better performance than using the LS-RED method in terms of (a) higher PSNR  and (b) faster convergence. 
This behavior is consistent with the results in \cite{Tirer2019BackProjectionBF}, which are not based on the RED approach.

\subsection{Super-resolution}\par

In the super-resolution (SR) task, the linear operator $\A$ represents filtering with an anti-aliasing kernel followed by subsampling.
We examine scale factors of 3 and 4. For scale factor of 3 we use a $7 \times 7$ Gaussian kernel with std of 1.6, and for scale factor of 4 we used a $9 \times 9$ Gaussian kernel with std of 2.2 (both are normalized to unit sum).


Here, $\A^\dagger$ in BP-RED is implemented using CG \cite{Hestenes&Stiefel:1952}, which has shown extremely fast convergence. Thus, again, the per-iteration complexity of LS-RED and BP-RED is dominated by the denoisers. 
We initialize both methods with bicubic upsampling of $\y$.
As before, we use the typical step-size of 1 over the Lipschitz constant of $\nabla \ell(\cdot)$ for both methods (here we do not examine the step-size from \cite{RED}, as it is not suitable for the noiseless case). 
For each method, the hyper-parameters 
$\lambda$ and $\sigma$ are tuned uniformly per scenario, using dense grids in the ranges of $(0.005,2.5)$ and $(0.5,20)$ respectively.

 

The restoration performance (average PSNR vs.~iteration number) of the two methods for all the experiments is presented in Fig.~\ref{fig:sr_iterations}.
The PSNR advantage of BP-RED over LS-RED is more moderate than in the deblurring case, but still clear and consistent.


\section{Conclusion}
\label{sec:con}

We considered the RED approach, which exploits existing image denoising engines for solving tasks other than denoising.
We showed that the RED gradient can be seen as a (sub)gradient of a prior function that is computed at a (typically slightly) denoised version of the point. 
This similarity between RED and traditional gradients motivated us to combine RED with the BP fidelity term, which has demonstrated improved results for traditional gradients schemes. 
Various experiments demonstrated the advantages of our BP-RED method over the combination of RED with least squares, used in previous works. 

\noindent {\bf Acknowledgment.}  This research was supported by ERC-StG grant no. 757497 (SPADE).


\bibliographystyle{IEEEbib}
\bibliography{refs}

\begin{thebibliography}{10}

\bibitem{donoho1994ideal}
D.~L. Donoho and J.~M. Johnstone,
\newblock ``Ideal spatial adaptation by wavelet shrinkage,''
\newblock {\em biometrika}, vol. 81, no. 3, pp. 425--455, 1994.

\bibitem{donoho1995noising}
D.~L. Donoho,
\newblock ``De-noising by soft-thresholding,''
\newblock {\em IEEE transactions on information theory}, vol. 41, no. 3, pp.
  613--627, 1995.

\bibitem{Rudin92TV}
L.~I. Rudin, S.~Osher, and E.~Fatemi,
\newblock ``Nonlinear total variation based noise removal algorithms,''
\newblock {\em Phys. D}, vol. 60, no. 1-4, pp. 259--268, Nov. 1992.

\bibitem{Dabov07BM3D}
K.~Dabov, A.~Foi, V.~Katkovnik, and K.~Egiazarian,
\newblock ``Image denoising by sparse 3-d transform-domain collaborative
  filtering,''
\newblock {\em IEEE Trans. on Image Processing}, vol. 16, no. 8, pp.
  2080--2095, 2007.

\bibitem{venkatakrishnan2013plug}
S.~V. Venkatakrishnan, C.~A. Bouman, and B.~Wohlberg,
\newblock ``Plug-and-play priors for model based reconstruction,''
\newblock in {\em 2013 IEEE Global Conference on Signal and Information
  Processing}. IEEE, 2013, pp. 945--948.

\bibitem{sreehari2016plug}
S.~Sreehari, S.~V. Venkatakrishnan, B.~Wohlberg, G.~T. Buzzard, L.~F. Drummy,
  J.~P. Simmons, and C.~A. Bouman,
\newblock ``Plug-and-play priors for bright field electron tomography and
  sparse interpolation,''
\newblock {\em IEEE Transactions on Computational Imaging}, vol. 2, no. 4, pp.
  408--423, 2016.

\bibitem{metzler2016denoising}
C.~A. Metzler, A.~Maleki, and R.~G. Baraniuk,
\newblock ``From denoising to compressed sensing,''
\newblock {\em IEEE Transactions on Information Theory}, vol. 62, no. 9, pp.
  5117--5144, 2016.

\bibitem{RED}
Y.~Romano, M.~Elad, and P.~Milanfar,
\newblock ``The little engine that could: Regularization by denoising (red),''
\newblock {\em SIAM Journal on Imaging Sciences}, vol. 10, no. 4, pp.
  1804--1844, 2017.

\bibitem{bigdeli2017deep}
S.~A. Bigdeli, M.~Zwicker, P.~Favaro, and M.~Jin,
\newblock ``Deep mean-shift priors for image restoration,''
\newblock in {\em Advances in Neural Information Processing Systems}, 2017, pp.
  763--772.

\bibitem{zhang2017learning}
K.~Zhang, W.~Zuo, S.~Gu, and L.~Zhang,
\newblock ``Learning deep cnn denoiser prior for image restoration,''
\newblock in {\em Proceedings of the IEEE conference on computer vision and
  pattern recognition}, 2017, pp. 3929--3938.

\bibitem{tirer2018image}
T.~Tirer and R.~Giryes,
\newblock ``Image restoration by iterative denoising and backward
  projections,''
\newblock {\em IEEE Transactions on Image Processing}, vol. 28, no. 3, pp.
  1220--1234, 2018.

\bibitem{buzzard2018plug}
G.~T. Buzzard, S.~H. Chan, S.~Sreehari, and C.~A. Bouman,
\newblock ``Plug-and-play unplugged: Optimization-free reconstruction using
  consensus equilibrium,''
\newblock {\em SIAM Journal on Imaging Sciences}, vol. 11, no. 3, pp.
  2001--2020, 2018.

\bibitem{reehorst2018regularization}
E.~T. Reehorst and P.~Schniter,
\newblock ``Regularization by denoising: Clarifications and new
  interpretations,''
\newblock {\em IEEE transactions on computational imaging}, vol. 5, no. 1, pp.
  52--67, 2018.

\bibitem{mataev2019deepred}
G.~Mataev, P.~Milanfar, and M.~Elad,
\newblock ``Deepred: Deep image prior powered by red,''
\newblock in {\em Proceedings of the IEEE International Conference on Computer
  Vision Workshops}, 2019, pp. 0--0.

\bibitem{sun2020block}
Y.~Sun, J.~Liu, and U.~S. Kamilov,
\newblock ``Block coordinate regularization by denoising,''
\newblock {\em IEEE Transactions on Computational Imaging}, vol. 6, pp.
  908--921, 2020.

\bibitem{sun2020async}
Y.~Sun, J.~Liu, Y.~Sun, B.~Wohlberg, and U.~S. Kamilov,
\newblock ``{Async-RED}: A provably convergent asynchronous block parallel
  stochastic method using deep denoising priors,''
\newblock {\em arXiv preprint arXiv:2010.01446}, 2020.

\bibitem{cohen2020regularization}
R.~Cohen, M.~Elad, and P.~Milanfar,
\newblock ``Regularization by denoising via fixed-point projection (red-pro),''
\newblock {\em arXiv preprint arXiv:2008.00226}, 2020.

\bibitem{Tirer2019BackProjectionBF}
T.~Tirer and R.~Giryes,
\newblock ``Back-projection based fidelity term for ill-posed linear inverse
  problems,''
\newblock {\em IEEE Transactions on Image Processing}, vol. 29, no. 1, pp.
  6164--6179, 2020.

\bibitem{tirer2019super}
T.~Tirer and R.~Giryes,
\newblock ``Super-resolution via image-adapted denoising cnns: Incorporating
  external and internal learning,''
\newblock {\em IEEE Signal Processing Letters}, vol. 26, no. 7, pp. 1080--1084,
  2019.

\bibitem{tirer2020convergence}
T.~Tirer and R.~Giryes,
\newblock ``On the convergence rate of projected gradient descent for a
  back-projection based objective,''
\newblock {\em arXiv preprint arXiv:2005.00959}, 2020.

\bibitem{zukerman2020bp}
J.~Zukerman, T.~Tirer, and R.~Giryes,
\newblock ``{BP-DIP}: A backprojection based deep image prior,''
\newblock {\em 2020 28th European Signal Processing Conference (EUSIPCO)}, pp.
  675--679, 2020.

\bibitem{Hestenes&Stiefel:1952}
M.~R. Hestenes and E.~Stiefel,
\newblock ``Methods of conjugate gradients for solving linear systems,''
\newblock {\em Journal of research of the National Bureau of Standards}, vol.
  49, pp. 409--436, 1952.

\end{thebibliography}

\end{document}